\documentclass[a4paper,11pt]{article}
\usepackage{pos}
\renewcommand{\logo}{\relax}
\usepackage{lineno}
\newcommand{\units}{\ \mathrm} 

\title{Probing new physics at the LUXE experiment}

\manuallySeparateAuthors
\author*[a]{Shan Huang}
\author{ for the LUXE Collaboration}
\affiliation[a]{School of Physics and Astronomy, Tel Aviv University,\\
  Tel Aviv-Yafo 6997801, Israel}
\emailAdd{shan.huang (at) desy.de}

\abstract{
The proposed LUXE experiment (Laser Und XFEL Experiment) at DESY Hamburg, using the electron beam from the European XFEL, aims to probe QED in the non-perturbative regime created in collisions between high-intensity laser pulses and high-energy electron or photon beams.
This setup also provides a unique opportunity to probe physics beyond the standard model.
In this proceedings, we show that by leveraging the large photon flux generated at LUXE, one can probe axion-like-particles (ALPs) up to a mass of 350 MeV and with a photon coupling of $3\times{10}^{-6}\ \mathrm{GeV}^{-1}$.
This reach is comparable to the background-free projection from NA62.
In addition, we will discuss other probes of new physics such as ALPs-electron coupling.
}

\FullConference{%
  41st International Conference on High Energy physics - ICHEP2022\\
  6-13 July, 2022\\
  Bologna, Italy
}

\begin{document}
\maketitle

\section{Introduction}
With the progress made in the past four decades, it is now possible to use laser to explore the frontier of particle physics \cite{DiPiazza_2012, Fedotov_2022}.
The LUXE, a proposed project combining a high-power laser and a high-energy electron beam from the European XFEL,
    has its main purpose to investigate the uncharted regime of non-perturbative QED \cite{LUXE_CDR_2021}.
The primary interaction in LUXE consists of an electron-photon or photon-photon collision at GeV scale.
The uniqueness of this experiment lies in its high-power laser, a 1.5-eV photon beam with high density, which comes with highly intense electromagnetic fields.
The strong fields also provide an opportunity to probe new physics in the intensity frontier, for instance, the existence of feebly-interacting particles \cite{Agrawal_2021} such as the axion-like particles (ALP).

Axions were proposed as a solution to the QCD strong-$CP$ problem \cite{Peccei_1977} and then became major candidates to the dark matter along with pseudoscalar or scalar ALPs \cite{Masso_2008}.
Other beyond-the-standard-model (BSM) candidates include the vector dark photon \cite{Holdom_1986}, the millicharged particle (mCP) \cite{Dobroliubov_1990}, and so on.
Nevertheless, no direct evidence of BSM particles has been found in any particle physics experiment.
In order to generate an observable signal, a large amount of photons are required in an experiment making use of the vertex between the BSM particle and the photon.
It is LUXE's capability to generate a high flux of GeV photons through an ``optical dump'', namely high-intensity Compton scattering (HICS) \cite{Bai_2021}.

In this proceedings, the methods of probing new physics at the optical dump (NPOD) in LUXE will be illustrated.
Shown in simulations, this light-shining-through-wall (LSW) type of ALP search is background free during the planned data taking after 2026.
With the 350 TW laser in Phase 1, the LUXE-NPOD bounds are comparable to the NA62-dump and the FASER2 experiments.
Additionally, the results of ALP-electron and mCP-photon couplings at the primary interaction point will also be presented.

\section{Methods}
One ALP couples to two photons in the minimal models.
The Lagrangian densities for ALPs of pseudoscalar ($a$) and scalar ($\phi$) fields are:
\begin{equation} \label{eq:agglagrangian}
    \mathcal{L}_{a,\phi} = \frac{a}{4\Lambda_a}\tilde{F}_{\mu\nu}F^{\mu\nu}+\frac{\phi}{4\Lambda_\phi}{F}_{\mu\nu}F^{\mu\nu},
\end{equation}
where $\Lambda_{a,\phi}$ are the coupling scales for $a\gamma\gamma$ and $\phi\gamma\gamma$ interactions, and $F$ and $\tilde{F}$ stand for the normal and dual electromagnetic field tensors.
The models also give the interaction widths $\Gamma_{X\gamma\gamma} = m_{X}^3/(64\pi\Lambda_{X}),$
in which $X$ stands for one of $a$ and $\phi$.
For reference, the ALP with $m_X=0.1\units{GeV}$ and $\Lambda_X=10^5\units{GeV}$ has a decay rate around $1\units{{ns}^{-1}}$ (in proper time).

\subsection{Optical dump}
\begin{figure}[htb]
    \centering
    \includegraphics[width=1\textwidth]{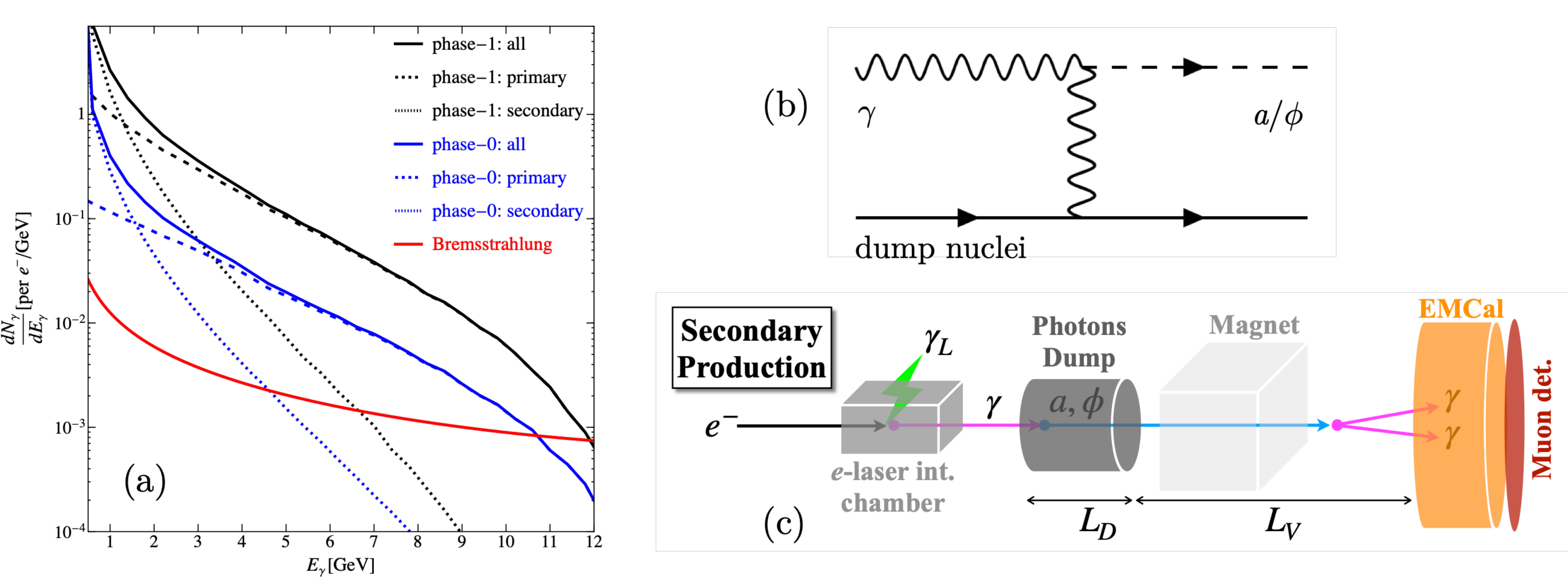}
    \caption{ \label{fig1}
        (a) Photon spectra of HICS by LUXE phase-0 and phase-1 lasers, and bremsstrahlung by tungsten target. The secondary productions are the photons created at the physical photon dump.
        (b) Feynman diagram for the related Primakoff process.
        (c) Sketch of LUXE-NPOD ALP production and detection scheme. \cite{Bai_2021}
    }
\end{figure}

In LUXE's electron-photon mode, a beam with $1.5\times 10^9$ 16.5-GeV electrons collide against an optical laser with intensity around $10^{18}$ to $10^{21}\units{W\ {cm}^{-2}}$.
This collision takes place in the strong-field regime and cannot be fully described by perturbative Compton scattering.
The theory of strong-field QED reveals the process of HICS having a different scattered photon spectrum and production.
Nevertheless, the time scale of this interaction stays at $\mathcal{O}(10)\units{fs}$, which is much shorter than the pair production of pico to nanoseconds.
It guarantees that the high flux of GeV photons remains a free stream, meaning the optical laser's electromagnetic field functions like a thick dump converting electrons into hard gamma ray.
Comparing to a ``real'' bremsstrahlung dump made of tungsten with $35\units{\mu m}$ in thickness, as Fig. \ref{fig1}(a) shows, the optical dump uplifts the hard photon production more than 20 times to $2.5\times 10^9$.
The laser of LUXE works at a 1 Hz repetition rate.
So a whole year run ($10^7$ bunch-crossings) will produce over $10^{16}$ hard photons for the study of new physics.

To reach the optimal laser configuration for Compton scattering, the laser's spatial-temporal parameters need to match the electron beam's characters.
However, this configuration may not be suitable for LUXE's primary goal to get as short and narrow as possible.
The aforementioned hard photon productions are calculated with a 40 W phase-0 laser, $6.5\units{\mu m}$ wide and 25 fs long 40-W, and 350-W phase-1 laser, $10\units{\mu m}$ wide and 120 fs long.

\subsection{Photon regeneration}
The existence of ALPs can be proven through photon regeneration \cite{Sikivie_1983, VanBibber_1987}.
The high flux of hard photons propagates along the beamline through photon detectors and hits onto a physical photon dump.
The dump will absorb the hard photons while converting a few of them into ALPs through the Primakoff process in the dump nucleus fields as in Fig. \ref{fig1}(b).
The ALP that lives in a suitable parameter space, if it exists, will decay into a couple of photons.
Since the HICS photons as well as other standard-model particles are blocked by the dump, a photon-pair signal picked up by the detector behind the dump will be the obvious and direct evidence for the ALP's existence.

\begin{figure}[htb]
    \centering
    \includegraphics[height=0.4\textwidth]{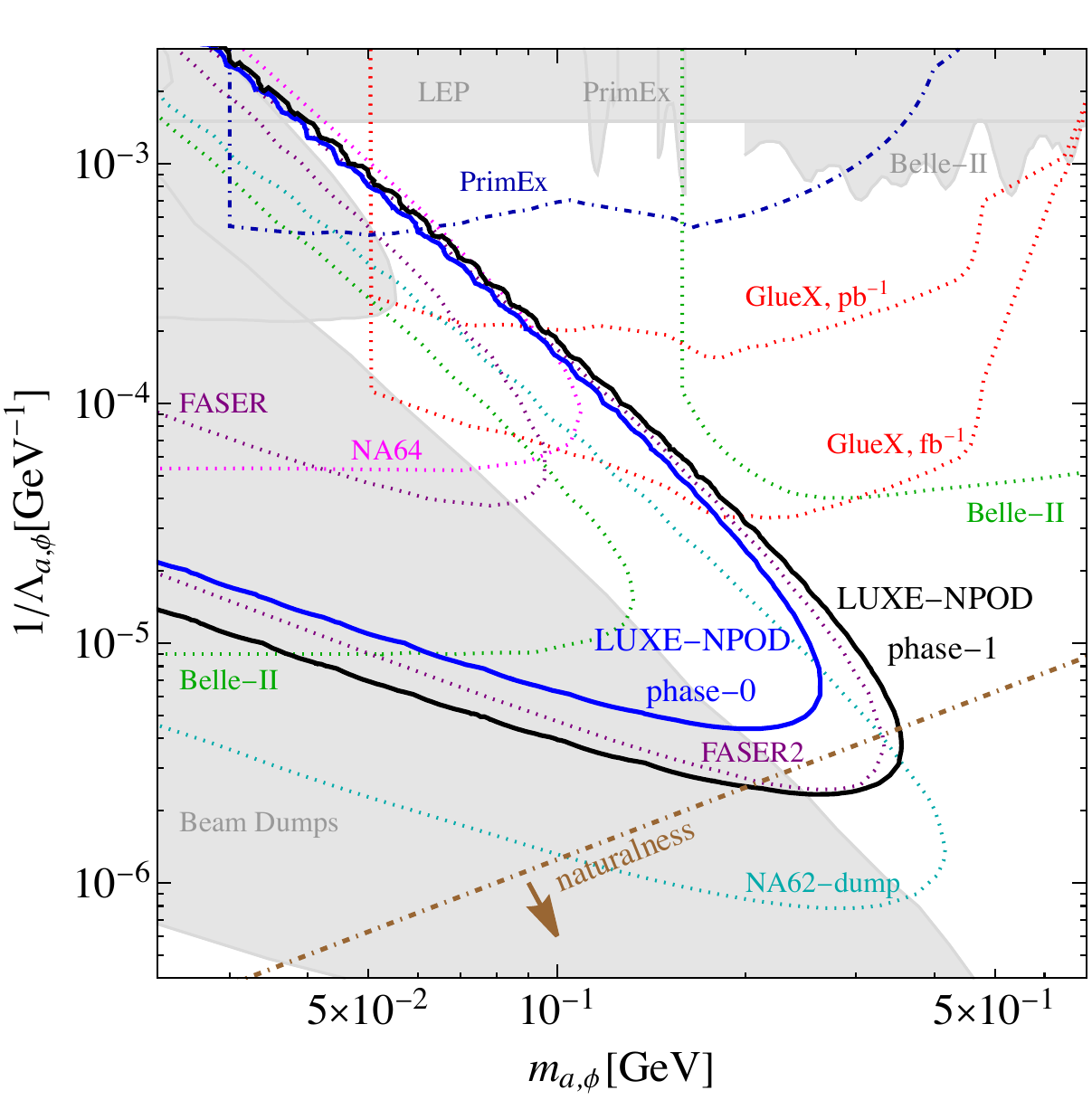}
    \hfill
    \includegraphics[height=0.38\textwidth]{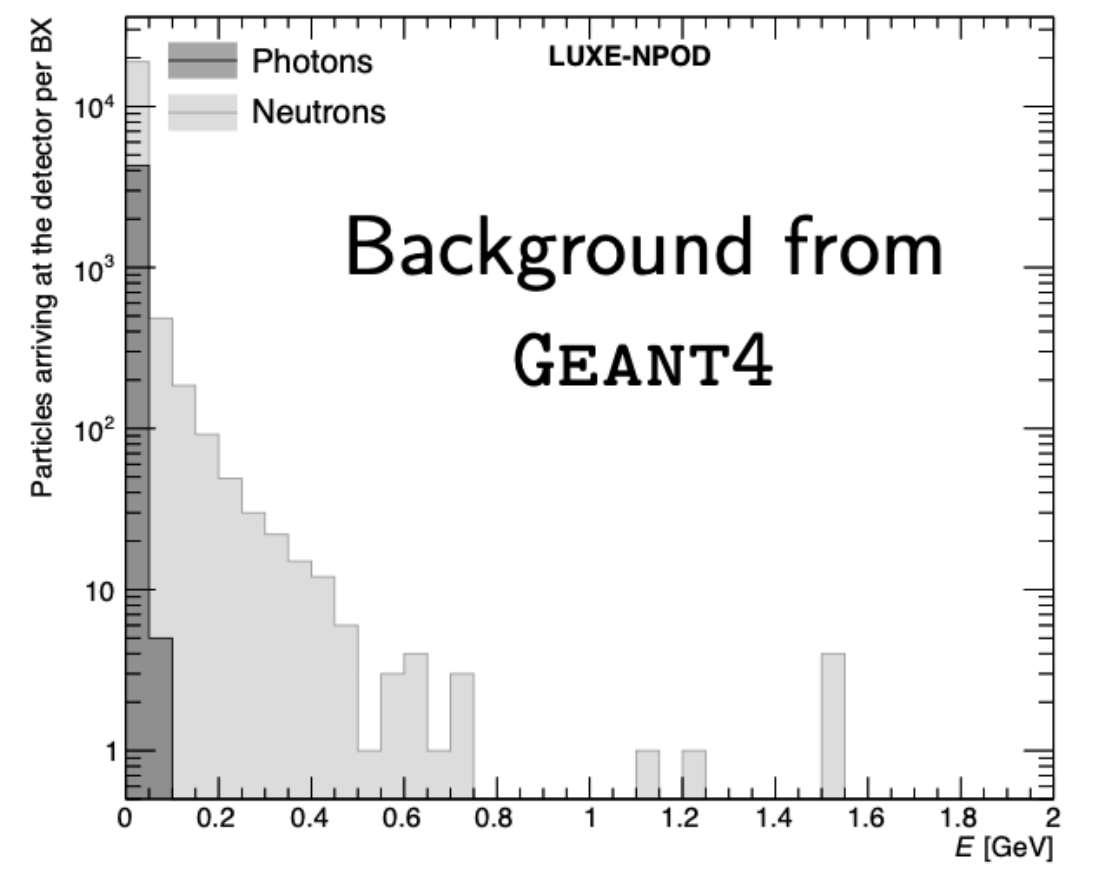}
    \caption{ \label{fig2}
        Left: Background-free LUXE-NPOD search bounds within similar experiments, with signal photon production $N_X \ge 3$.
        Right: Background photon and neutron spectra per bunch-crossing. \cite{Bai_2021}
    }
\end{figure}

The production of signal pairs depends on the effective luminosity $\mathfrak{L}_\mathrm{eff}$ of the hard photon flux,
    the hard photon spectrum $\mathrm{d}N_\gamma / \mathrm{d}E_\gamma$,
    the cross section $\sigma_{X}(E_\gamma)$ of the $X\gamma\gamma$ process,
    the detector acceptance $\mathcal{A}$,
    and the geometry [as in Fig. \ref{fig1}(c)].
It can be estimated as
\begin{equation} \label{eq:alpproduction}
    N_{X} \approx \mathfrak{L}_\mathrm{eff} \int{
        \mathrm{d} E_\gamma \frac{\mathrm{d}N_\gamma}{\mathrm{d}E_\gamma} \sigma_{X}(E_\gamma)
        \left(\exp{\left\{-\frac{L_\mathrm{D}}{L_\mathrm{X}}\right\}} - \exp{\left\{-\frac{L_\mathrm{V}+L_\mathrm{D}}{L_\mathrm{X}}\right\}} \right) \mathcal{A}}.
\end{equation}
The luminosity relates not only to the multiplicity of hard photons but also to the material of the physical dump.
Figure \ref{fig1}(a) shows the photon spectrum.
The acceptance relies on the transverse size of the detector and its efficiency.
The geometry determines the detectable ALP parameters.
A longer dump will absorb signal photon pairs regenerated from shorter-lived ALPs, while a short distance will also prevent the ALPs decaying within $L_\mathrm{D}+L_\mathrm{V}$.
Hence, a smaller $L_\mathrm{D}$ and larger $L_\mathrm{V}$ will benefit the search bounds.
Nonetheless, these two parameters are constrained by the dump effectiveness and the spatial limit in the experimental hall.
The LUXE search bounds with $L_\mathrm{D}=1\units{m}$ and $L_\mathrm{V}=2.5\units{m}$ for $N_X \ge 3$ in a one-year background-free run are shown in Fig. \ref{fig2}(left).
The estimation in Eq.~(\ref{eq:alpproduction}) agrees with the \textsc{MadGraph} Monte Carlo results.

\subsection{Background estimation}
Although a one-meter dump can block most of the standard-model particles, there are still some leaks from multiple scattering inside the dump.
The background particles consist of charged particles, neutrons, and hadronic photons.
Hence as in Fig. \ref{fig1}(c), a magnet dipole is put between the dump and the detector to bend away charged particles.
A \textsc{Geant4} simulation shows that the background has about 10 neutrons and $10^{-2}$ photons per bunch-crossing [Fig. \ref{fig2}(right)].
The time of arrival is used to cut out the slow-flying neutrons, and the signal is formed by a pair of photons arriving at the same time.
With 0.5 ns time rejection and diphoton matching event reconstruction, the background rate is reduced to $<10^{-7}\units{Hz}$, marking LUXE-NPOD a background-free run.

\section{BSM study at the IP}
\begin{figure}[htb]
    \centering
    \includegraphics[width=0.65\textwidth]{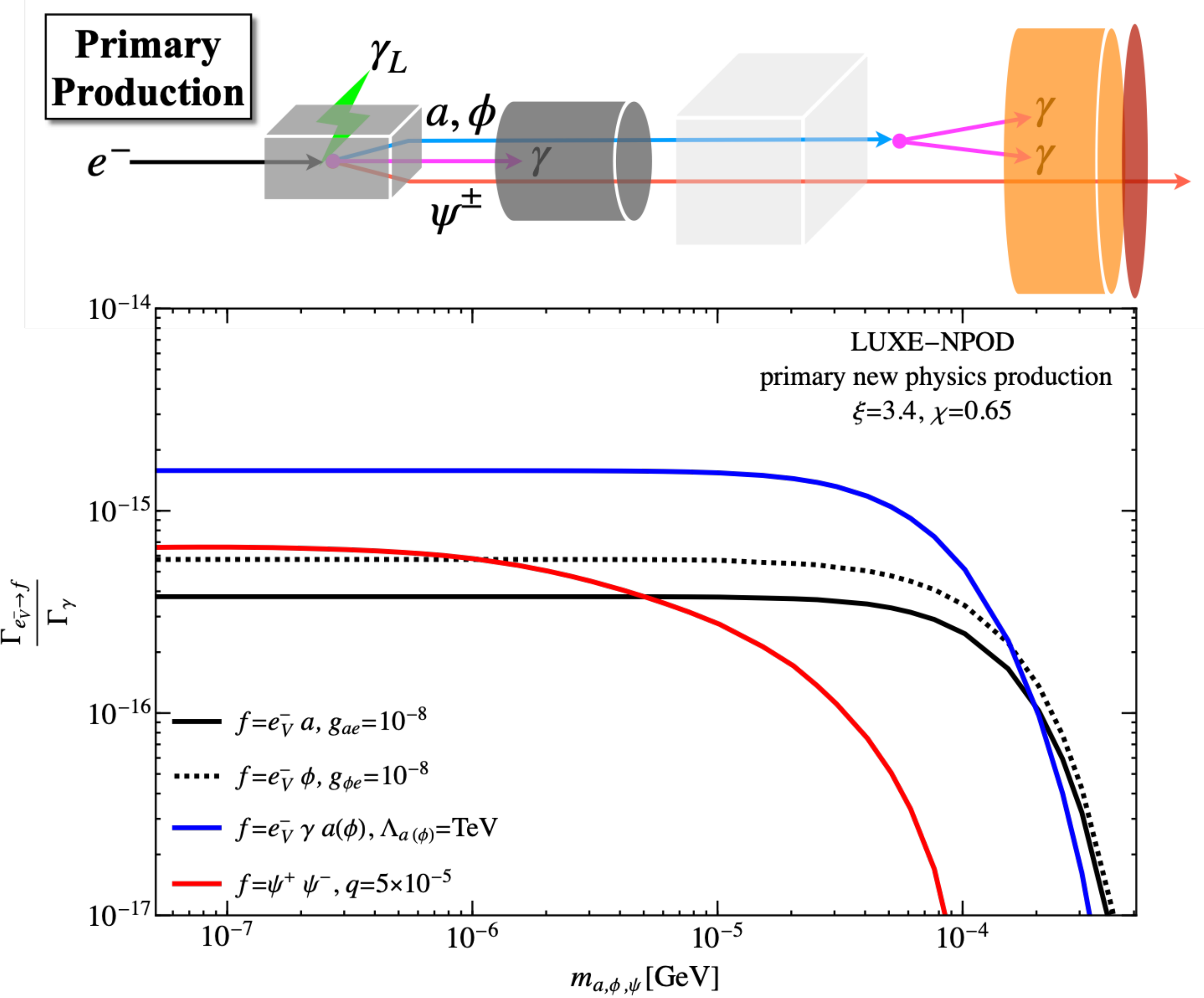}
    \caption{ \label{fig3}
        Top: Sketch of ALP or mCP production at LUXE interaction point.
        Bottom: ALP and mCP production efficiencies for given coupling coefficients and laser intensity parameter $\xi$. \cite{Bai_2021}
    }
\end{figure}

The pseudoscalar and scalar ALPs can also couple to electrons as
\begin{equation} \label{eq:aeelagrangian}
    \mathcal{L}_{ae,\phi e} = i g_{ae} a \bar{e} \gamma^5 e + g_{\phi e} \phi \bar{e} e,
\end{equation}
where $e, \bar{e}$ are electron spinor fields, and $g_{ae,\phi e}$ are the coupling coefficients.
And for another BSM candidate the millicharged particle with spinor field $\psi$ and charge $q \ll q_e$, its interaction with electrons and photons has a Lagrangian density
$    \mathcal{L}_{\psi} = eq\bar{\psi}\gamma_\mu A^\mu \psi.$
It is possible for these particles to decay back to the standard-model particles within $L_\mathrm{V}$ as in Fig. \ref{fig3}.
Although calculations under the assumption of constant laser field at the IP roughly showed that the production rates are much lower than the LUXE-NPOD $X\gamma\gamma$ process for the given BSM particle parameters.

\section{Discussion and Conclusion}
In LUXE, the high-intensity Compton scattering of 16.5-GeV electrons by an optical dump of the laser strong fields generates a free-stream and bright GeV photon flux.
The LUXE-NPOD is planned to use this flux exploring the proposed interaction between photons and axion-like-particles through a background-free LSW-type experiment.
Estimates have shown that the LUXE-NPOD scheme can reach the uncharted ALP parameter space for ALP-photon interactions.
In addition, the high flux of hard photons also provides unique opportunities for other open questions.
For example, aside from ALPs and mCPs, the X17 particle, related to an anomaly at 17 MeV \cite{Wong_2020}, could be produced during the interaction between the hard photons and the dump.
And in photonuclear physics, the accurate study on the reverse oxygen-carbon reaction in the triple alpha process \cite{Gai_2020} demands a gamma ray of huge amount of photons with a narrow energy spectrum.
Generally, the high-quality and high-multiplicity gamma ray source by HICS has outstanding potential in experimental physics.
And the LUXE-NPOD experiment will start its data taking after 2026.

\section*{Acknowledgements}
This work was partly supported by the German-Israeli Foundation and the PAZY Foundation.
I would like to thank my LUXE colleagues for their help in producing this write-up.
We thank the DESY technical staff for continuous assistance and the DESY directorate for their strong support and the hospitality they extend to the non-DESY members of the collaboration.
This work has benefited from computing services provided by the German National Analysis Facility (NAF) and the Swedish National Infrastructure for Computing (SNIC).
I would also like to thank Prof. Cheuk-Yin Wong from ORNL for his comments and the inspiring discussions.

\bibliographystyle{JHEP}
\bibliography{luxenpod_ichep}

\end{document}